\documentclass[twocolumn]{jpsj3} 

\usepackage{graphicx, color}

\title{Ground-State Phase Diagram of the Multiple-Spin Exchange Model with up to the Six-Spin Exchange Interactions on the Triangular Lattice}

\author{Chitoshi Yasuda$^{1}$\thanks{E-mail address: cyasuda@sci.u-ryukyu.ac.jp}, Yuma Uchihira$^{2}$, Sh\^ogo Taira$^{3}$, and Kenn Kubo$^{2}$} 

\inst{$^{1}$Department of Physics and Earth Sciences, University of the Ryukyus, Nishihara, Okinawa 903-0213, Japan \\
$^{2}$Department of Physics and Mathematics, Aoyama Gakuin University, Fuchinobe, Sagamihara 252-5258, Japan \\
$^{3}$Graduate School of Engineering and Science, University of the Ryukyus, Nishihara, Okinawa 903-0213, Japan
}

\abst{The ground state of the multiple-spin exchange model with up to the six-spin exchange interactions on a triangular lattice in the magnetic field is investigated within the mean-field approximation. By comparing the phase diagrams of systems with and without the five- and six-spin interactions, we find that the five- and six-spin interactions expand the ferromagnetic phase. In contrast, the antiferromagnetic-like phases are contracted. A phase with a 1/2 plateau is also contracted by the five- and six-spin interactions. In addition to the results that have not yet been reported for systems with up to the four-spin interactions, we obtain new phases for systems with up to the six-spin interactions. In particular, a 1/6 plateau appears in the low magnetic field near the ferromagnetic phase in the magnetization process. This phase has a twelve-sublattice structure in which a unit cell consists of seven spins parallel to the magnetic field and five spins antiparallel to it. 
}

\recdate{\today}

\begin{document}
\maketitle

\section{Introduction}

Two-dimensional (2D) solid helium has attracted considerable attention owing to the expectation of the appearance of novel quantum phases and quantum many-body phenomena induced by strong quantum fluctuations and frustrations~\cite{Fukuyama2008}. At low millikelvin temperatures, helium atoms adsorbed on the surface of graphite form a thin film consisting of a monolayer, two layers, or three layers depending on the density $\rho$ of the atoms. The atoms in the second layer of two or three layers can move relatively freely because the base graphite has a negligible effect on the second layer. Thus, an ideal 2D spin system is realized.~\cite{Godfrin1988} In the 4/7 commensurate phase stabilized at $\rho \simeq 6.4$ nm$^{-2}$, the helium atoms form a triangular lattice~\cite{Greywall1990}, and  the system is regarded as a spin system with the multiple-spin exchange interactions due to ring  exchanges of the $^3$He nuclei.~\cite{Delrieu1980, Roger1984} For $^3$He films with $\rho \simeq 7.85$ nm$^{-2}$, the three-spin exchange interactions are larger than other interactions, e.g., the two- and four-spin exchange interactions.~\cite{Bernu1992} Furthermore, a comparison of experimental results for heat capacity and magnetic susceptibility with those of exact diagonalization in the multiple-spin exchange model suggests that the five- and six-spin exchange interactions have a nonnegligible effect.~\cite{Roger1990} 

In 2D solid $^3$He, the two-spin exchange frequency is large for a sufficiently low density, and as the density is increased, the two-spin exchange frequency decreases more rapidly than the three-spin exchange frequency.~\cite{Roger1990,Roger1983} Experimental measurements of heat capacity and magnetic susceptibility have also shown that the solid $^3$He layer changes from an antiferromagnetic system in which the two-spin exchange is dominant to a ferromagnetic system in which the three-spin exchange is dominant depending on the coverage.~\cite{Siqueira1996} These results indicate that interactions with various values of the parameters are realized in the solid $^3$He layer. 

Furthermore, multiple-spin exchanges reportedly play an important role in the inorganic compound $\kappa$-(ET)$_2$Cu$_2$(CN)$_3$.~\cite{Motrunich2005} Because this compound is considered to be realized in a parameter region different from that of the solid $^3$He layer, we need to investigate the ground states of the multiple-spin exchange model for various parameter regions.  

Because the multiple-spin exchange model on the triangular lattice describes a quantum spin system with strong frustration, there is no efficient theoretical method except for the spin-wave theory and the exact diagonalization of a small system. In this work, we investigate the ground state of the multiple-spin exchange model within the mean-field approximation to advance quantitative understanding. The classical ground states can be used to examine the effect of quantum fluctuations by using the spin-wave theory~\cite{Taira2018}.  In the model with up to the four-spin exchange interactions, the ground-state phase diagram has already been investigated within the mean-field approximation.~\cite{Kubo1997, Momoi1999, Kubo2003} The details of the results will be mentioned in Sect.~3. In this work, we reveal the effects of the five- and six-spin exchange interactions on the ground state by investigating the model with up to the six-spin exchange interactions.

The remainder of this article is organized as follows. We introduce the Hamiltonian in Sect.~2 and explain the ground-state phase diagram in the absence of the five- and six-spin exchange interactions in Sect.~3. In Sect.~4, we compare the ground-state phase diagrams of the system with and without the five- and six-spin exchange interactions. Two new phases are induced by the five- and six-spin exchange interactions. Finally, we present a summary in Sect.~5.

\section{Model Hamiltonian with up to the Six-Spin Exchange Interactions}

The Hamiltonian with two-, three-, four-, five-, and six-spin exchange interactions on a triangular lattice is given by 
\begin{equation}
 {\cal H}=\sum_{n=2}^6 {\cal H}_n \ ,
\end{equation}
with
\begin{equation}
 {\cal H}_n=(-1)^{n} J_n \sum_{n\mbox{-}{\rm spin~ring}} (P_n + P_n^{-1}) \ ,
\end{equation}
where $J_n$ represents the positive exchange constants of $n$-spin ring exchanges, and $P_n$ and $P_n^{-1}$ are the $n$-spin ring exchange operators and their inverse operators, respectively~\cite{Thouless1965}. Because odd and even particle exchange interactions are ferromagnetic and antiferromagnetic, respectively, the system creates frustrations due to the competing multiple-spin exchange interactions. 

The two-spin Hamiltonian ${\cal H}_2$ is written using the Pauli matrix $\mib{\sigma}_i$ associated with a spin-1/2 particle at site $i$ (hereafter, $\mib{\sigma}_i$ is called spin), 
\begin{equation}
 {\cal H}_2=  \frac{J_2}{2}  \sum_{\rm bond} (1 + \mib{\sigma}_1 \cdot \mib{\sigma}_2) \ ,
\end{equation}
where $\sum_{\rm bond}$ is the summation over all $3N$ bonds on the triangular lattice, and $\mib{\sigma}_1$ and $\mib{\sigma}_2$ are the spins at two edge sites on a bond, with $N$ being the number of sites. The three-spin Hamiltonian is also written using  the two-spin interactions, i.e.,
\begin{equation}
 {\cal H}_3 = -\frac{J_3}{2} \sum_{\rm triangle} (1 + \mib{\sigma}_1 \cdot
  \mib{\sigma}_2 + \mib{\sigma}_2 \cdot \mib{\sigma}_3 + \mib{\sigma}_3
  \cdot \mib{\sigma}_1) \ ,
\end{equation}
where $\sum_{\rm triangle}$ is the summation over all $2N$ triangles, and $\mib{\sigma}_1$, $\mib{\sigma}_2$, and $\mib{\sigma}_3$ are the spins at three vertex sites on a triangle.
\begin{figure}[t]
\begin{center}
\includegraphics[width=0.4\textwidth]{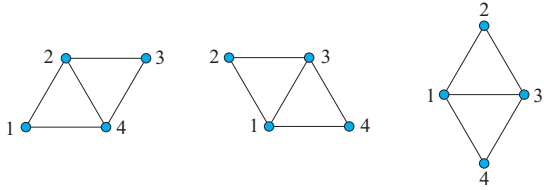}
\caption{Illustration of three types of plaquettes for the four-spin interactions.}
\label{fig-4spin}
\end{center}
\end{figure}
The four-spin Hamiltonian is written as
\begin{eqnarray}
&& \hspace*{-1cm} {\cal H}_4 = \frac{J_4}{4} \sum_{\rm plaq} \{1+ \sum_{1 \le \alpha <
  \beta \le 4} \mib{\sigma}_{\alpha} \cdot \mib{\sigma}_{\beta} +
  (\mib{\sigma}_1 \cdot \mib{\sigma}_2)(\mib{\sigma}_3 \cdot
  \mib{\sigma}_4) \nonumber \\
  && + (\mib{\sigma}_1 \cdot \mib{\sigma}_4)(\mib{\sigma}_2
  \cdot \mib{\sigma}_3) - (\mib{\sigma}_1 \cdot
  \mib{\sigma}_3)(\mib{\sigma}_2 \cdot \mib{\sigma}_4) 
  \} \ ,
  \label{eq:H4}
\end{eqnarray}
where $\sum_{\rm plaq}$ is the summation over all $3N$ plaquettes, and $\mib{\sigma}_1$, $\mib{\sigma}_2$, $\mib{\sigma}_3$, and $\mib{\sigma}_4$ are the spins at four vertex sites on a plaquette.  There are three types of plaquettes as shown in Fig.~\ref{fig-4spin}.
\begin{figure}[t]
\begin{center}
\includegraphics[width=0.4\textwidth]{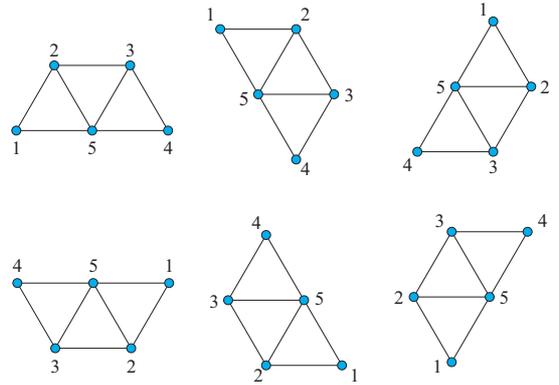}
\caption{Illustration of six types of trapezoids for the five-spin interactions.}
\label{fig-5spin1}
\end{center}
\end{figure}
\begin{figure}[t]
\begin{center}
\includegraphics[width=0.5\textwidth]{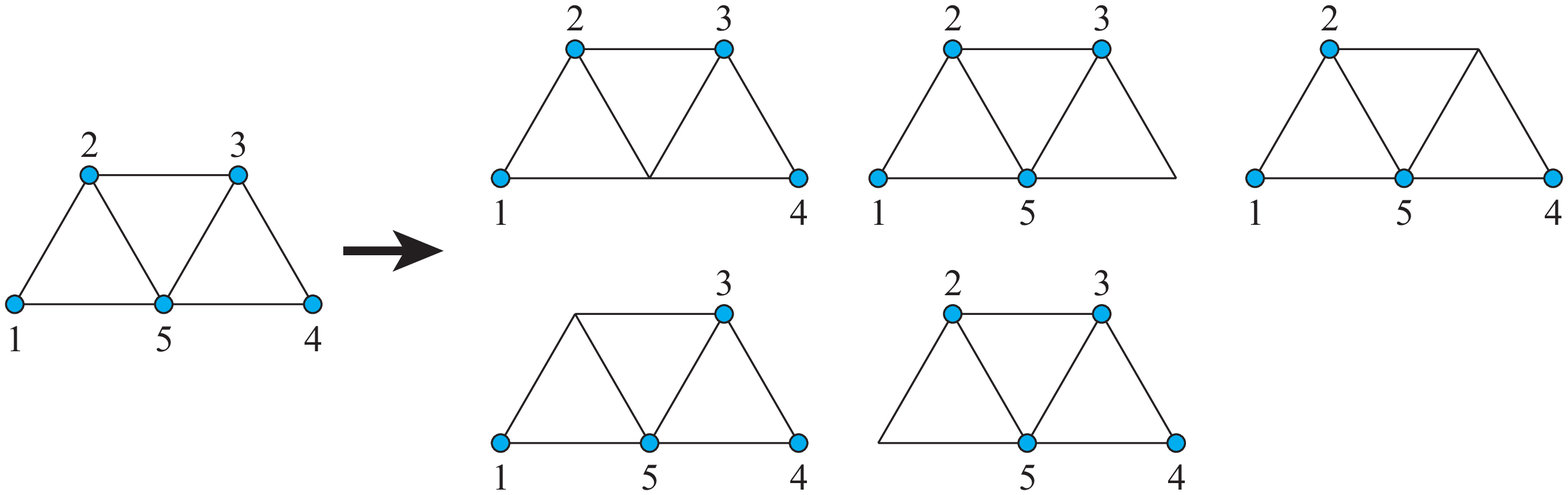}
\caption{All combinations of four spins in a trapezoid. The four solid circles for each trapezoid describe four spins.}
\label{fig-5spin2}
\end{center}
\end{figure}
The five-spin Hamiltonian is written as
\begin{eqnarray}
\label{5spin}
&& \hspace*{-0.5cm} {\cal H}_5 = - \frac{J_5}{8} \sum_{\rm trap} [1 + \sum_{1 \leq \alpha < \beta \leq
 5} \mib{\sigma}_{\alpha} \cdot \mib{\sigma}_{\beta} \nonumber \\
 && + \sum_{l=1}^{5} \{(\mib{\sigma}_{\alpha_l} \cdot \mib{\sigma}_{\beta_l})(\mib{\sigma}_{\gamma_l} \cdot \mib{\sigma}_{\delta_l})  + (\mib{\sigma}_{\beta_l} \cdot \mib{\sigma}_{\gamma_l})(\mib{\sigma}_{\delta_l} \cdot \mib{\sigma}_{\alpha_l}) \nonumber \\
 &&- (\mib{\sigma}_{\alpha_l} \cdot
  \mib{\sigma}_{\gamma_l})(\mib{\sigma}_{\beta_l} \cdot
  \mib{\sigma}_{\delta_l}) \}] \ , 
\end{eqnarray}
where $\sum_{\rm trap}$ is the summation over all $6N$ trapezoids and $\alpha_l=l$, $\beta_l={\rm mod}(l,5)+1$, $\gamma_l={\rm mod}(l+1,5)+1$, $\delta_l={\rm mod}(l+2,5)+1$. There are six types of trapezoids as shown in Fig.~\ref{fig-5spin1}. The summation $\sum_{l=1}^5$ in Eq.~(\ref{5spin}) takes over all combinations of four spins in a trapezoid as shown in Fig.~\ref{fig-5spin2}. 
\begin{figure}[t]
\begin{center}
\includegraphics[width=0.5\textwidth]{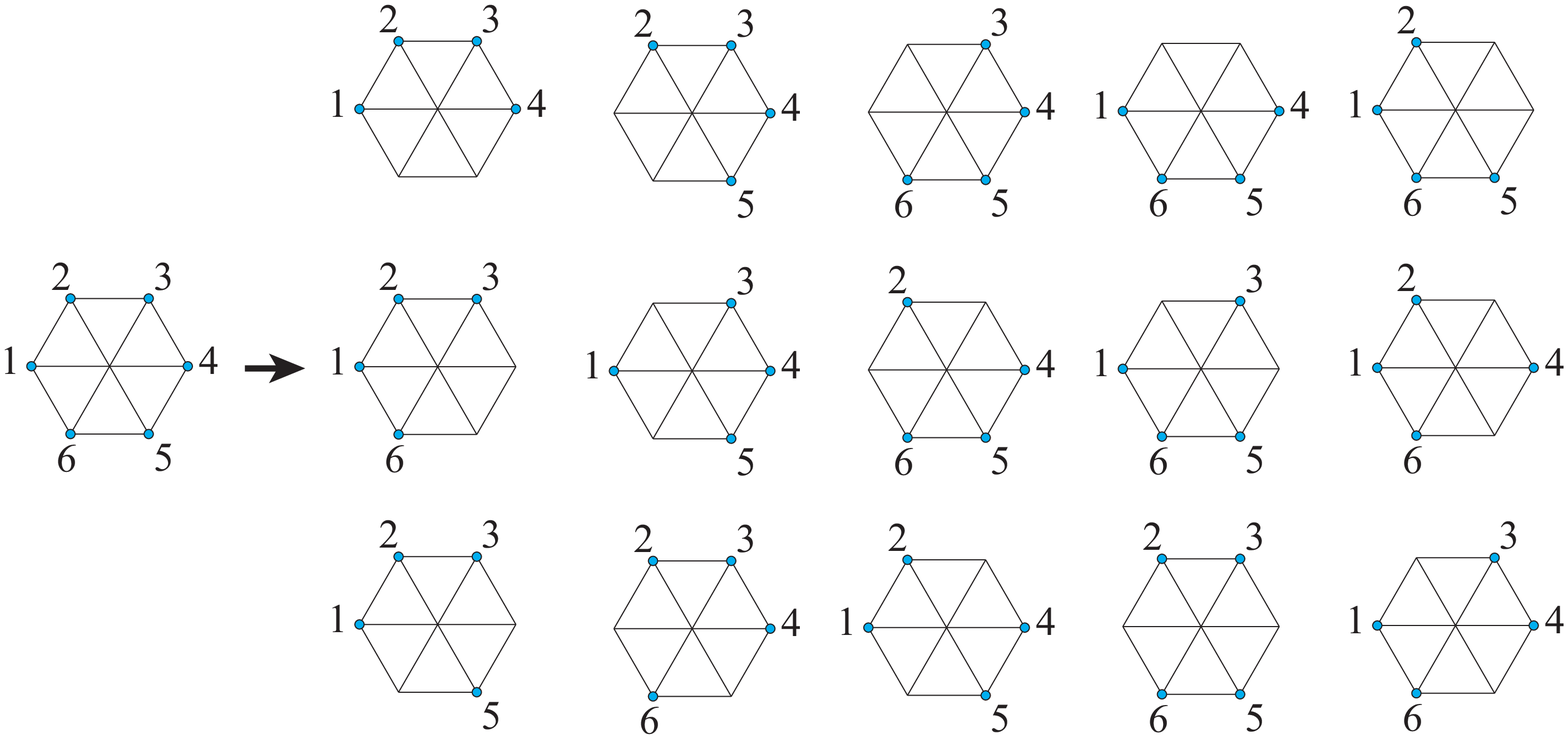}
\caption{Illustration of the six-spin interactions and all combinations of four spins in a hexagon.}
\label{fig-6spin1}
\end{center}
\end{figure}
\begin{figure}[t]
\begin{center}
\includegraphics[width=0.45\textwidth]{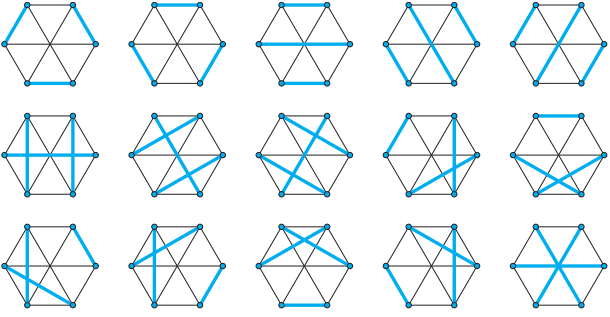}
\caption{All combinations of six spins in a hexagon. The six solid circles included in three solid bonds describe six spins.}
\label{fig-6spin2}
\end{center}
\end{figure}
The six-spin Hamiltonian is written as
\begin{eqnarray}
 && {\cal H}_{6} = \frac{J_6}{16} \sum_{\rm hexa} [1 + \sum_{1 \leq \alpha < \beta \leq 6} \mib{\sigma}_{\alpha} \cdot \mib{\sigma}_{\beta} \nonumber \\
&&+ \sum_{l=1}^{6} \{(\mib{\sigma}_{\alpha_l} \cdot
 \mib{\sigma}_{\beta_l})(\mib{\sigma}_{\gamma_l} \cdot
 \mib{\sigma}_{\delta_l})  + (\mib{\sigma}_{\beta_l} \cdot
 \mib{\sigma}_{\gamma_l})(\mib{\sigma}_{\delta_l} \cdot
 \mib{\sigma}_{\alpha_l}) \nonumber \\
&&- (\mib{\sigma}_{\alpha_l} \cdot
 \mib{\sigma}_{\gamma_l})(\mib{\sigma}_{\beta_l} \cdot
 \mib{\sigma}_{\delta_l}) \} \nonumber \\
&&+ \sum_{l=1}^{6} \{(\mib{\sigma}_{\alpha_l} \cdot
 \mib{\sigma}_{\beta_l})(\mib{\sigma}_{\gamma_l} \cdot
 \mib{\sigma}_{\zeta_l})  + (\mib{\sigma}_{\beta_l} \cdot
 \mib{\sigma}_{\gamma_l})(\mib{\sigma}_{\zeta_l} \cdot
 \mib{\sigma}_{\alpha_l}) \nonumber \\
 &&- (\mib{\sigma}_{\alpha_l} \cdot
 \mib{\sigma}_{\gamma_l})(\mib{\sigma}_{\beta_l} \cdot
 \mib{\sigma}_{\zeta_l}) \} \nonumber \\
&&+ \sum_{l=1}^{3} \{(\mib{\sigma}_{\alpha_l} \cdot
 \mib{\sigma}_{\beta_l})(\mib{\sigma}_{\delta_l} \cdot
 \mib{\sigma}_{\zeta_l})  + (\mib{\sigma}_{\beta_l} \cdot
 \mib{\sigma}_{\delta_l})(\mib{\sigma}_{\zeta_l} \cdot
 \mib{\sigma}_{\alpha_l}) \nonumber \\
 &&- (\mib{\sigma}_{\alpha_l} \cdot
 \mib{\sigma}_{\delta_l})(\mib{\sigma}_{\beta_l} \cdot
 \mib{\sigma}_{\zeta_l}) \} \nonumber \\
&&+ \sum_{l=1}^2  (\mib{\sigma}_{\alpha_l} \cdot
 \mib{\sigma}_{\beta_l})(\mib{\sigma}_{\gamma_l} \cdot
 \mib{\sigma}_{\delta_l})(\mib{\sigma}_{\zeta_l} \cdot
 \mib{\sigma}_{\kappa_l}) \nonumber \\ 
 &&+ \sum_{l=1}^3  (\mib{\sigma}_{\alpha_l} \cdot
 \mib{\sigma}_{\delta_l})(\mib{\sigma}_{\beta_l} \cdot
 \mib{\sigma}_{\gamma_l})(\mib{\sigma}_{\zeta_l} \cdot
 \mib{\sigma}_{\kappa_l}) \nonumber \\ 
&&+ \sum_{l=1}^3  (\mib{\sigma}_{\alpha_l} \cdot
 \mib{\sigma}_{\delta_l})(\mib{\sigma}_{\gamma_l} \cdot
 \mib{\sigma}_{\zeta_l})(\mib{\sigma}_{\beta_l} \cdot
 \mib{\sigma}_{\kappa_l}) \nonumber \\
 &&- \sum_{l=1}^6  (\mib{\sigma}_{\alpha_l} \cdot
 \mib{\sigma}_{\beta_l})(\mib{\sigma}_{\gamma_l} \cdot
 \mib{\sigma}_{\zeta_l})(\mib{\sigma}_{\delta_l} \cdot
 \mib{\sigma}_{\kappa_l}) \nonumber \\ 
&&- (\mib{\sigma}_1 \cdot \mib{\sigma}_4)(\mib{\sigma}_2 \cdot
 \mib{\sigma}_5)(\mib{\sigma}_3 \cdot \mib{\sigma_6}) ] \ , \label{6spin}
\end{eqnarray}
where $\sum_{\rm hexa}$ is the summation over all $N$ hexagons, and $\alpha_l=l$, $\beta_l={\rm mod}(l,6)+1$, $\gamma_l={\rm mod}(l+1,6)+1$, $\delta_l={\rm mod}(l+2,6)+1$, $\zeta_l={\rm mod}(l+3,6)+1$, $\kappa_l={\rm mod}(l+4,6)+1$. We adopt only a regular hexagon as the six-spin cyclic exchange because the occurrence probability of the other pattern of the six-spin exchange is negligible owing to the effects of interatomic potentials in solid $^3$He.~\cite{Ceperley1987, Bernu1992} The summation of the products of four spins in Eq.~(\ref{6spin}) is taken over all combinations of four spins in a hexagon as shown in Fig.~\ref{fig-6spin1}. The summation of the products of six spins in Eq.~(\ref{6spin}) is taken over all combinations of six spins in a hexagon as shown in Fig.~\ref{fig-6spin2}.

Neglecting the constant terms, applying the transformations ${\cal H}_4=\frac{J_4}{4} \sum_{\rm plaq} h_4$, ${\cal H}_5=-\frac{J_5}{8} \sum_{\rm trap} h_5$, ${\cal H}_6=\frac{J_6}{16} \sum_{\rm hexa} h_6$, $J=\frac{J_2}{2}-J_3$, $K=\frac{J_4}{4}$, $L=-\frac{J_5}{8}$, $M=\frac{J_6}{16}$, and adding the Zeeman term with the external magnetic field $\mib{H}$, we obtain the Hamiltonian
\begin{eqnarray}
 && \hspace{-1cm} {\cal H} = J \sum_{<i,j>}\mib{\sigma}_i \cdot \mib{\sigma}_j + K
  \sum_{\rm plaq} h_4 + L \sum_{\rm trap} h_5 + M \sum_{\rm hexa} h_6
  \nonumber \\
  &&- \sum_i \mib{H} \cdot \mib{\sigma}_i \ ,
  \label{hamiltonian}
\end{eqnarray}
 where $\sum_{<i,j>}$ is the summation over all nearest-neighbor pairs. The direction of the magnetic field $\mib{H}$ is regarded as the $z$-direction. Although $K$ and $M$ are always positive and $L$ is negative, $J$ has either sign. Because the magnitude of $J_2$ is expected to be larger than $J_3$ at a low atomic density, the parameter $J$ is expected to be positive at a low density and negative at a high density.
 
From comparison of the experimental results and theoretical results of the path integral Monte Carlo method, exact diagonalization, and high-temperature expansion for specific heat and magnetic susceptibility, the values of $J/J_4$, $J_5/J_4$, and $J_6/J_4$ for the $^3$He layer have been evaluated~\cite{Roger1998, Bernu1992, Misguich1998, Masutomi2004}. In these results, $J/J_4$ is approximately $-0.45$ or $-1$; i.e., $J/K=-1.8$ or $-4$. Although the situation in which $^3$He atoms are adsorbed is different for each study, $J_5/J_4$ is approximately 0.3 for all the studies. On the other hand,  $J_6/J_4$ varies from 0.08 to 1. This variation of $J_6/J_4$ might be caused by the sensitivity of $J_6/J_4$ to the density of the $^3$He atoms. In this work, we investigate the system with $J_5/J_4=J_6/J_4=0.3$, i.e., $L/K=-0.15$ and $M/K=0.075$ mainly on the basis of these previous results. In addition, we estimate the phase diagram of the system with $J_5/J_4=0.3$ and $J_6/J_4=0.6$, i.e., $L/K=-0.15$ and $M/K=0.15$ to discuss the influence of $J_6/J_4$.

\section{Phase Diagram of the System with up to  the Four-Spin Exchange Interactions}
 
In this work, we investigate the effects of the five- and six-spin exchange interactions on the ground state within the mean-field approximation assuming $6\times6$ sublattices. We neglect quantum fluctuations and treat $\mib{\sigma}_i$ as a unit vector. Thus, the problem becomes that of minimalization of the classical ground-state energy. In this section, we show the ground-state phase diagram in the absence of the five- and six-spin interactions, i.e., $L=M=0$, in order to compare the phase diagrams with and without the five- and six-spin interactions. Although many results on the phase diagram have already been published~\cite{Kubo1997, Momoi1999, Kubo2003},  we show the phase diagram in wider parameter ranges and the order of phase transitions. Furthermore, we obtain new results that have not been reported in previous works. We also explain the properties of each phase for later discussion. The phase-transition points are estimated by the conjugate gradient (CG) method. In this method, we prepare approximately $10^2 - 10^4$ random spin states as the initial states. The number of random states depends on the parameters of the system that we calculate. In addition to the random states, we prepare some expected states as the initial states. We can find unexpected phase transitions by the CG method. Various physical quantities are used to estimate the phase transitions, e.g., energy, magnetization, and vector chirality, which is defined as
\begin{equation}
 \mib{\chi}^{\rm v}
 = \frac{1}{N} \sum_{\triangle} (\mib{\sigma}_1 \times \mib{\sigma}_2 + \mib{\sigma}_2 \times \mib{\sigma}_3 + \mib{\sigma}_3 \times \mib{\sigma}_1) \ ,
 \label{vec_chiral}
\end{equation}
and the scalar chiralities, which are defined as
\begin{equation}
  \chi^{\rm s} (\mib{Q}) = \frac{1}{N} \sum_{\triangle i}~e^{-i \mib{Q} \cdot \mib{r}_i}~\mib{\sigma}_1 \cdot (\mib{\sigma}_2 \times \mib{\sigma}_3) \ ,
 \label{sca_chiral}
\end{equation}
with the wave number $\mib{Q}$, where $\mib{\sigma}_1$, $\mib{\sigma}_2$, and $\mib{\sigma}_3$ are the spins on the vertices of a triangle. The summations in Eqs.~(\ref{vec_chiral}) and (\ref{sca_chiral}) are taken over all upward triangles.  

\begin{figure}[t]
\centering
\includegraphics[width=0.45\textwidth]{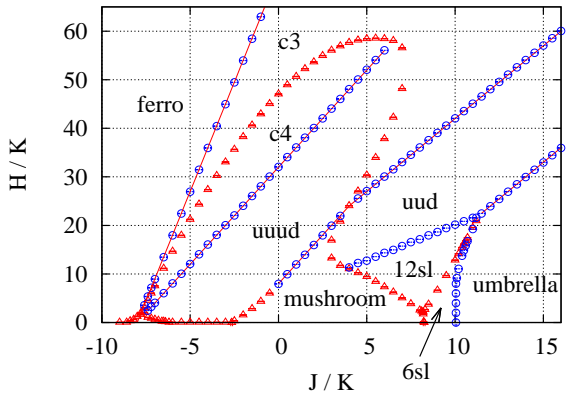}
  \caption{Ground-state phase diagram of the system with up to the four-spin exchange interactions ($L$=$M$=0) calculated assuming $6\times6$ sites. The horizontal and vertical axes are the magnitudes of the two-spin interactions and magnetic field divided by that of the four-spin interactions, respectively.  The 
triangles and circles with error bars show the first- and second-order phase transitions estimated by the CG method, respectively. The lines are the phase-transition lines obtained within some assumptions.}
\label{36site-phase}
\end{figure}

\subsection{Phase diagram and phases}

We show the ground-state phase diagram parametrized by $J/K$ and $H/K$ in Fig.~\ref{36site-phase}, where $H$ is the magnitude of the magnetic field. Nine main phases appear, and each phase is labeled in Fig.~\ref{36site-phase}. First, we explain each phase.

\begin{figure}[t]
\centering
\includegraphics[width=0.45\textwidth]{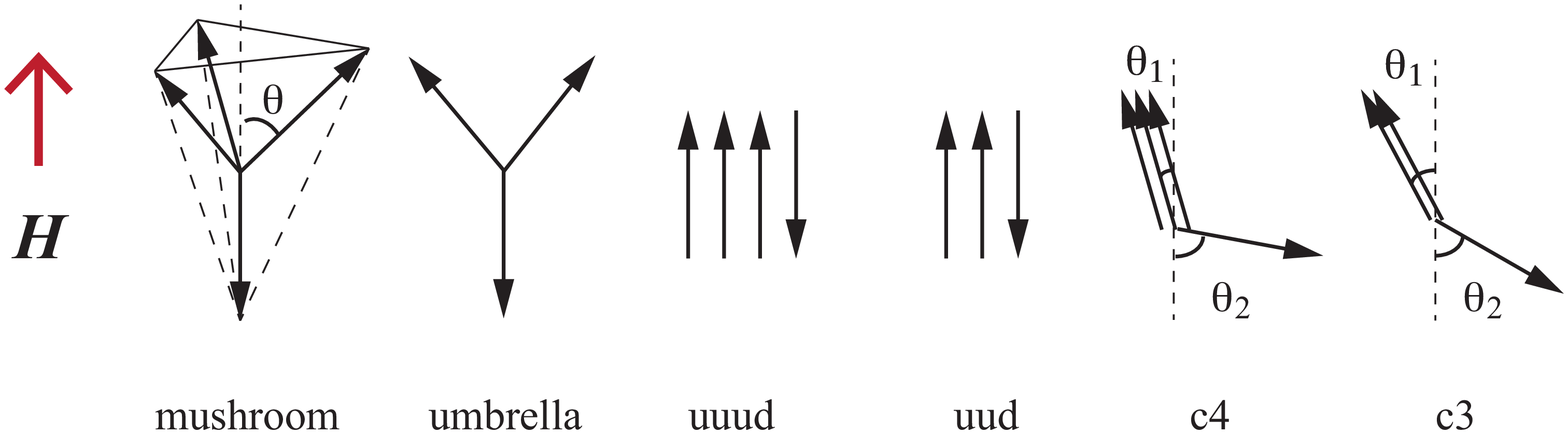}
  \caption{Spin states appearing as the ground state for the multiple-spin exchange model in the magnetic field within the mean-field approximation. The direction of the magnetic field $\mib{H}$ is also described by the thick arrow.}
\label{phases}
\end{figure}

\begin{enumerate}
 \item ferro: The ground state is the ferromagnetic state with full magnetization.
 \item mushroom: Spins on one sublattice point downward, and those on the other three sublattices are oriented upward but canted by an angle $\theta$ with respect to the field as shown in Fig.~\ref{phases}. There is threefold rotational symmetry about the field axis. The ground state has a finite value of the uniform scalar chirality $\chi^{\rm s}(\mib{0})$. 
 \item umbrella: The ground state has a coplanar spin configuration where the spins on one sublattice point downward, and those on the other two sublattices are oriented upward but canted by an angle with respect to the field. There is a finite vector chirality $\mib{\chi}^{\rm v}$. 
 \item uuud: The collinear spins on three sublattices point upward, whereas those on the other sublattice point downward. The ground state realizes a 1/2 plateau in the magnetization process.
 \item uud: The ground state has collinear spins similar to those of the uuud state and realizes a 1/3 plateau in the magnetization process.
 \item c4: The ground state has a four-sublattice structure. The spins on three sublattices are parallel to each other, and all the spins are canted with respect to the magnetic field. The two angles of the spin vectors satisfy the relation $\sin{\theta_2}=3\sin{\theta_1}$, where $\theta_1$ is the angle from the $z$-axis of the three parallel spins, and $\theta_2$ is the angle from the $z$-axis of the other spin, as shown in Fig.~\ref{phases}.
 \item c3: The ground state has coplanar spins similar to those of the c4 state. The two angles of the spin vectors satisfy the relation $\sin{\theta_2}=2\sin{\theta_1}$.
 \item 6sl: The ground state has a non-coplanar spin structure on six sublattices with the uniform vector chiral order $\mib{\chi}^{\rm v}$ and the staggered scalar chiral order $\chi^{\rm s}(\mib{Q})$ with $\mib{Q}=(\pi,\pi)$, $(\pi,0)$, or $(0,\pi)$. The details of this state have already been investigated.~\cite{Yasuda2007}
 \item 12sl: The ground state has a twelve-sublattice structure. The details of this state will be explained in Subsect.~3.3.
\end{enumerate}

In addition to the phases mentioned above, small phases with various structures appear in the low-field region $-9 < J/K < -2.6$. Because the detailed classification is beyond the scope of this work, we do not precisely estimate these phases. In a previous work~\cite{Kubo2003} assuming 144 sublattices,  more states were found in the region $-8.61 < J/K < -2.26$. These small phases might be artifacts of the assumption of a finite number of sublattices. The true ground state in this parameter region is still controversial, even at the mean-field level. 

The triangles and circles in Fig.~\ref{36site-phase} indicate the first- and second-order phase-transition points estimated by the CG method, respectively. The error bars correspond to the intervals between the calculated data points. The lines in Fig.~\ref{36site-phase} are the phase-transition lines analytically obtained by assuming the two phases and the order of the phase transition. We list the concrete expression of the critical magnetic field $H_{\rm c}$ in Table~\ref{hc}. Note that these functions are the results for the system including the five- and six-spin interactions. The critical magnetic field $H_{\rm c}$ between the ferro and uuud phases is derived from the crossing point of the ground-state energy curves assuming the first-order phase transition. The other $H_{\rm c}$ values in Table~\ref{hc} are derived assuming the second-order phase transition. For $L=M=0$, the phase transition between the ferromagnetic and uuud phases is not realized. The results described by the solid lines in Fig.~\ref{36site-phase} agree with those obtained by the CG method. In this work, we found that the phase transition between the mushroom and uuud phases changes from the second order to the first order near $J/K=0$. We explain the details in Subsect.~3.2.

\begin{table}[tb]
\caption{Critical magnetic field $H_{\rm c}$ on the phase transition between phases 1 and 2. The phase transition between the ferro and uuud phases is evaluated assuming the first-order phase transition, and the others are evaluated assuming the second-order phase transition.}
\begin{center}
\begin{tabular}{lll}
   Phase 1 & Phase 2 & \hspace{1cm}$H_{\rm c}$ \\ \hline
   umbrella & uud & $3(J-4K+8L+32M)$ \\
   uud & c3 & $3(J+4K+8L)$ \\
   ferro & c3 & $9(J+8K+40L+16M)$ \\
   mushroom & uuud & $4(J+2K+12M)$ \\
   uuud & c4 & $4(J+8K+24L)$ \\ 
   ferro & uuud & $6(J+8K+32L+8M)$ \\ \hline
\end{tabular}
\end{center}
\label{hc}
\end{table}

The phase-transition points in Fig.~\ref{36site-phase} are evaluated by the CG method as follows. Points on the ferro, uuud, uud, c3, and c4 phases are evaluated mainly by the magnetization process. Points on the mushroom, 6sl, and umbrella phases are evaluated mainly using the scalar chirality $\chi^{\rm s}(\mib{0})$, the staggered scalar chirality $\chi^{\rm s}(\mib{Q})$ with $\mib{Q}=(\pi, \pi)$, $(\pi, 0)$, and $(0, \pi)$, and the vector chirality $\mib{\chi}^{\rm v}$, respectively. We also confirmed the ground-state energy for all phase transitions. The exact confirmation of the order of  the phase transitions by numerical methods is difficult. Note that we confirm the type of order within the interval between the calculated data points.

Next, we explain the phase diagram in Fig.~\ref{36site-phase} itself. For $|J|/K \gg 1$ and the ferromagnetic $J < 0$, the ferromagnetic state is stabilized. For $|J|/K \gg 1$ and the antiferromagnetic $J > 0$, the state with the umbrella structure is stabilized. Because the model is regarded as the Heisenberg model on the triangular lattice in the region where $J$ is dominant, it is easy to understand that these states are stabilized. For $|J|/K \ll 1$, the state with the mushroom structure is stable owing to the antiferromagnetic four-spin interactions. Although both the two- and four-spin interactions are antiferromagnetic, they compete because the two- and four-spin interactions stabilize the three- and four-sublattice structures, respectively. Also, the magnetic field competes with interactions. As a result of these competitions, the uuud, c3, and c4 phases appear between the ferromagnetic and mushroom phases, and the phases with twelve- and six-sublattice structures appear between the umbrella and mushroom phases.

With increasing field, the system undergoes a phase transition from the umbrella, 6sl, 12sl, and mushroom phases to the uud and uuud phases. One of the interesting behaviors is the appearance of a magnetization plateau with 1/2 of the full polarization in the parameter region suitable for $^3$He layers~\cite{Kubo1997, Misguich1998, Momoi1999, Nema2009}. This plateau is realized by the uuud structure. As the field increases further, the system undergoes a phase transition from the uud and uuud phases to the phases with the c3 and c4 structures. For the phase transition between the umbrella and 6sl phases, the boundary was previously determined by the spin-wave theory, and the results agree with those of the present work.~\cite{Yasuda2006}

\begin{figure}[t]
\centering
	{\resizebox{0.45\textwidth}{!}{\includegraphics{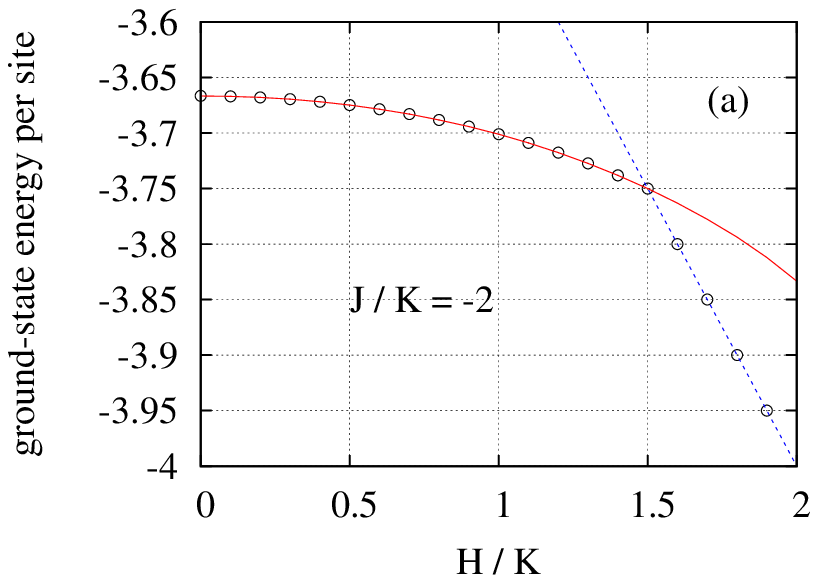}}} 	
	{\resizebox{0.45\textwidth}{!}{\includegraphics{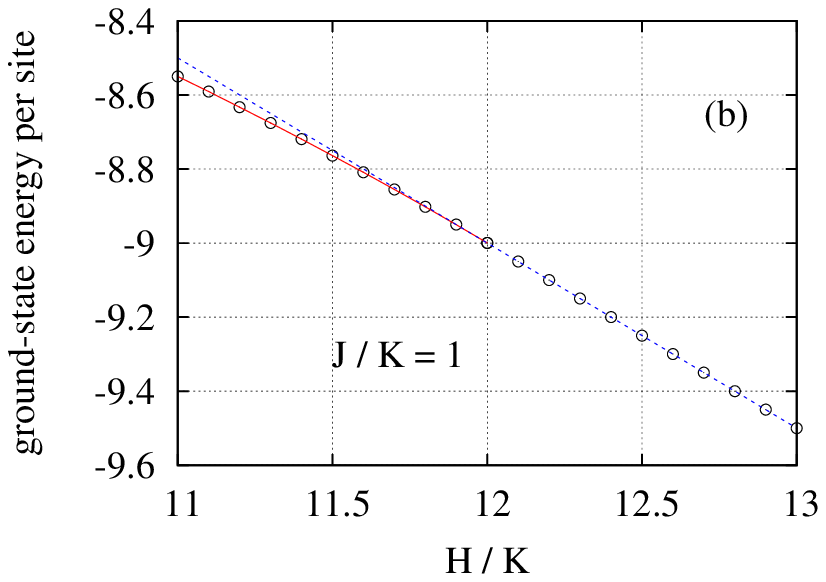}}}
  \caption{Dependences of the ground-state energy per site on the magnetic field $H/K$ for (a) $J/K = -2$  and (b) 1. The circles are the results calculated by the CG method. The solid and broken lines are the results analytically calculated assuming the mushroom and uuud structures, respectively.}  
\label{mush-uuud}
\end{figure}

\subsection{Phase transition between mushroom and uuud phases}

As mentioned above, the phase transition between the mushroom and uuud phases changes from second order to first order near $J/K=0$. In Fig.~\ref{mush-uuud}, we show the $H/K$ dependence of the ground-state energy for $J/K = -2$ and 1. The circles are the results calculated by the CG method. The solid and broken lines are the results calculated analytically assuming the mushroom and uuud structures, respectively. The agreement between these results shows that the phase transition is that between the mushroom and uuud phases. While the ground-state energies of the two states intersect for $J/K = -2$,  the energy of the mushroom structure connects to that of the uuud structure smoothly for $J/K = 1$. For $J/K>0$, the angle $\theta$ by which three spins are canted with respect to the magnetic field continuously approaches zero. On the other hand, for $J/K<0$, before the angle reaches zero, the phase transition occurs.

\subsection{Twelve-sublattice structure}

\begin{figure}[t]
\centering
\includegraphics[width=0.35\textwidth]{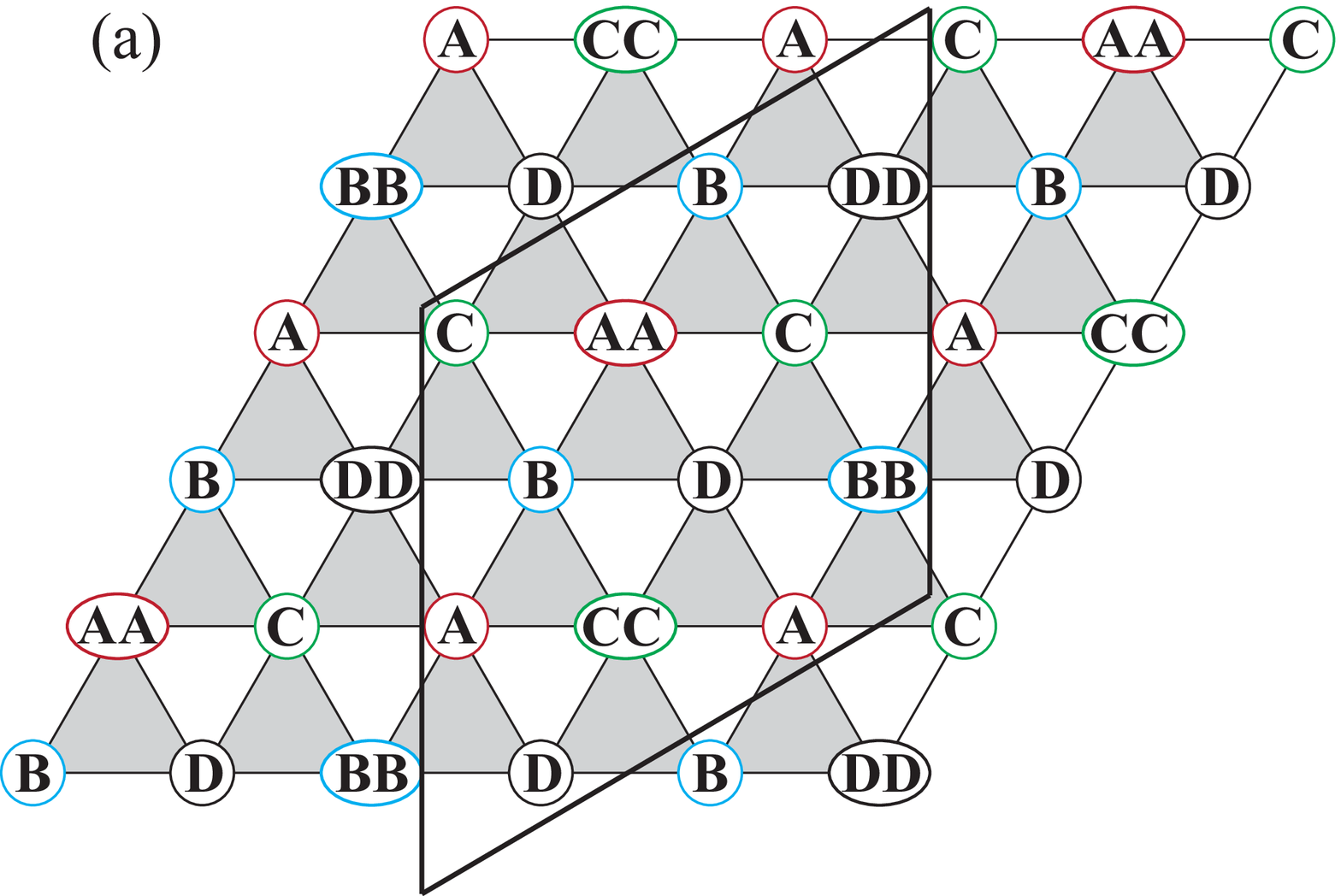}
\includegraphics[width=0.35\textwidth]{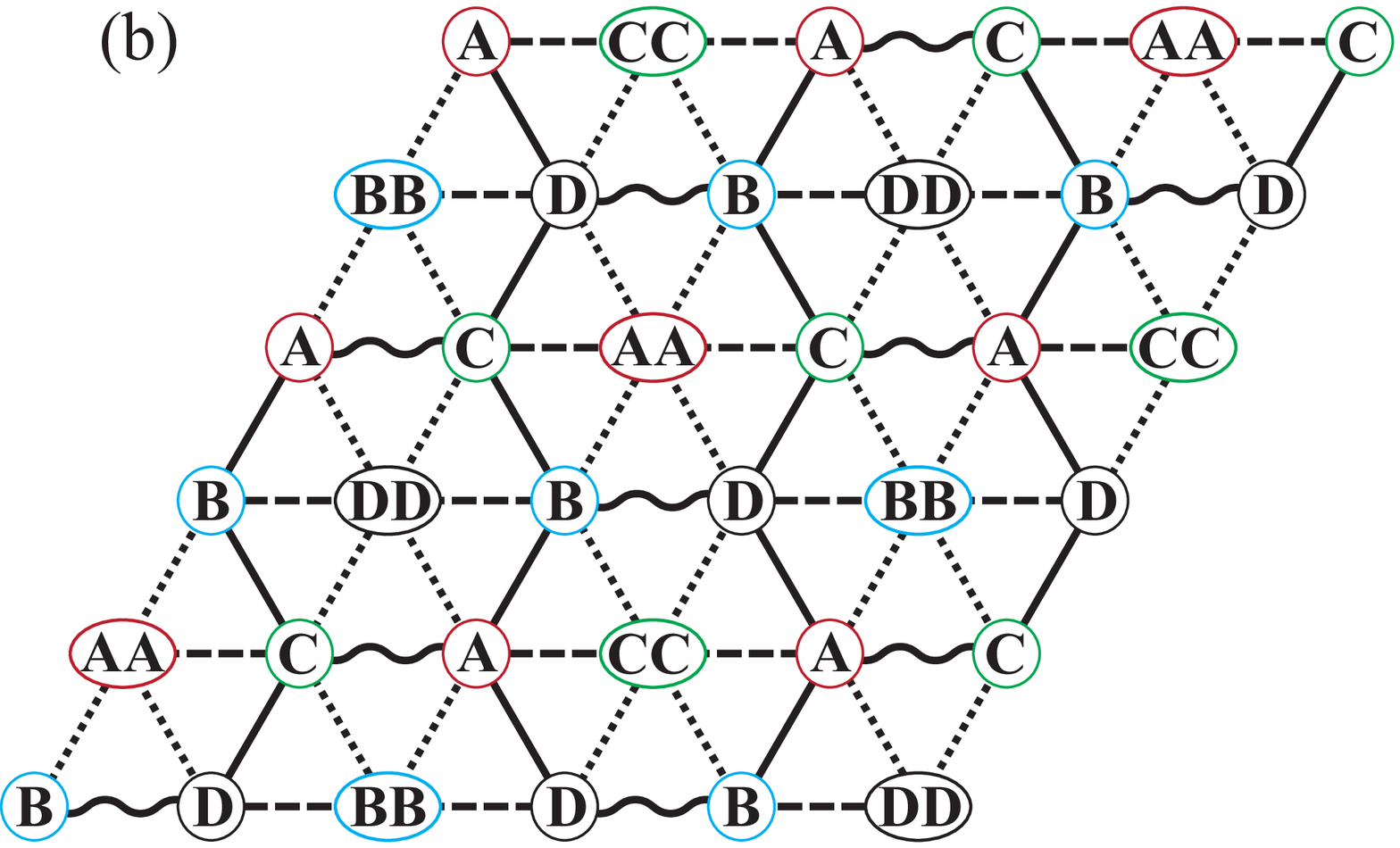}
  \caption{(a) Example of the spin configuration of the twelve-sublattice structure and (b) the correlations between the nearest-neighbor spins. Although the unit cell surrounded by the solid line in the upper panel consists of twelve spins, the number of types of spin orientation is eight. The marks A, B, C, D, AA, BB, CC, and DD describe the types of spin orientation. The solid, wavy, dotted, and broken bonds in the lower panel denote the nearest-neighbor correlation functions with the strengths of approximately 0.65, 0.29, $-0.77$, and $-0.95$, respectively.}
\label{12sl-phase}
\end{figure}

\begin{figure}[t]
\centering
  {\resizebox{0.4\textwidth}{!}{\includegraphics{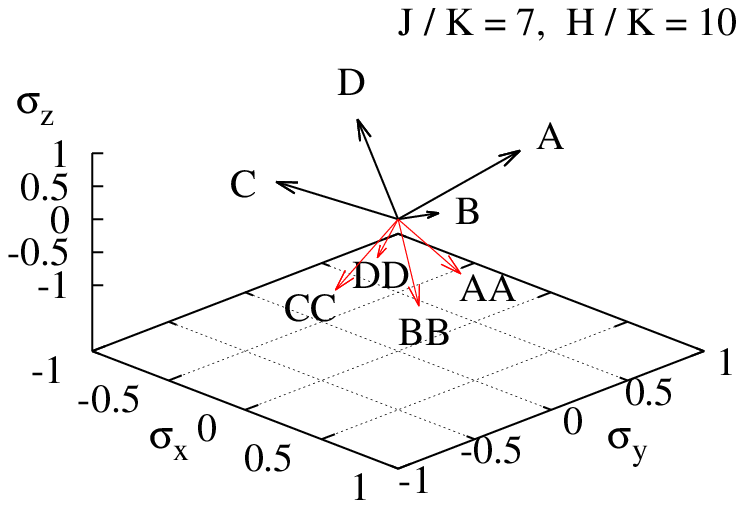}}}
  {\resizebox{0.4\textwidth}{!}{\includegraphics{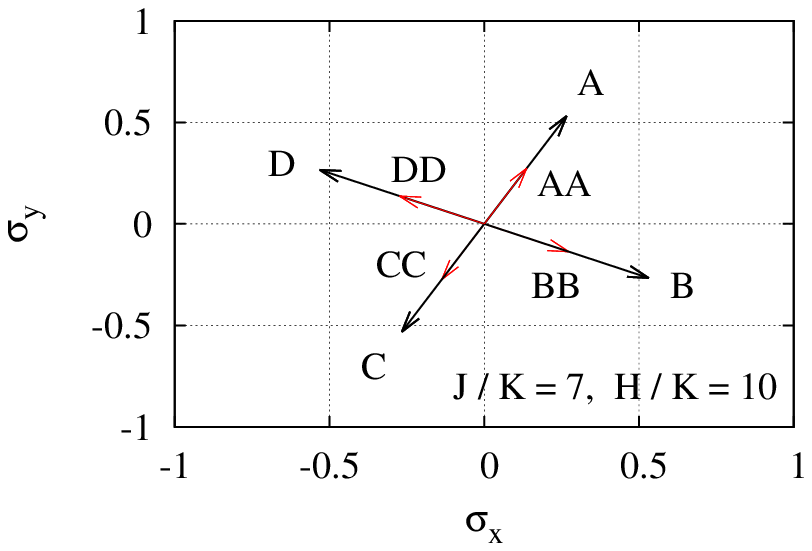}}}
  \caption{Eight types of spin orientation and the $x$ and $y$ components constructing the twelve-sublattice structure for $J/K=7$ and $H/K=10$. The magnitude of the spin vectors is fixed to unity.}
\label{12sl-8spins}
\end{figure}

The phase with the twelve-sublattice (12sl) structure is surrounded by the phases with mushroom, 6sl, and uud structures, as shown in Fig.~\ref{36site-phase}. Although the structure of the 6sl phase has already been investigated within the mean-field approximation~\cite{Yasuda2007}, that of the 12sl phase has not yet been investigated. The 12sl structure connects the three-, four-, and six-sublattice structures. The unit cell of the 12sl structure is shown in Fig.~\ref{12sl-phase}(a). Although the unit cell surrounded by the solid line consists of twelve spins, the spin orientation is classified into eight types. In Fig.~\ref{12sl-phase}, the types are labeled A, B, C, D, AA, BB, CC, and DD. A unit cell consists of two A, B, C, and D spins and single AA, BB, CC, and DD spins as shown in Fig.~\ref{12sl-phase}(a). As shown in Fig.~\ref{12sl-8spins}, we find that  A and AA are in the same plane perpendicular to the $xy$ plane and that their $z$ components have opposite signs. The same situation holds for B and BB, C and CC, and D and DD. 

We show the nearest-neighbor correlation functions $\langle \mib{\sigma}_i \cdot \mib{\sigma}_j \rangle$ in Fig.~\ref{12sl-phase}(b). The solid, wavy, dotted, and broken bonds in Fig.~\ref{12sl-phase}(b) denote $\langle \mib{\sigma}_i \cdot \mib{\sigma}_j \rangle$ with the strengths of approximately 0.65, 0.29, $-0.77$, and $-0.95$, respectively, which are calculated for $J/K = 7$ and $H/K = 10$. This system has four four-spin correlation functions defined by $\langle h_4 \rangle$, where $h_4$ is the four-spin operator that we obtain by neglecting the constant terms of ${\cal H}_4$ expressed by Eq.~(\ref{eq:H4}) and multiplying the result by $\frac{4}{J_4}$. Classifying the correlation of each plaquette according to the diagonal bond, we obtain $\langle h_4 \rangle$ of the plaquettes with the solid, wavy, dotted, and broken diagonal bonds evaluated as approximately $-0.98$, $-1.02$, $-1.17$, and $-1.60$, respectively. 
In the umbrella state for $J/K = 15$ and $H/K = 10$, there are two types of $\langle \mib{\sigma}_i \cdot \mib{\sigma}_j \rangle$ with the strengths of approximately $-0.63$ and $-0.22$, and there are two corresponding types of $\langle h_4 \rangle$ with the strengths of approximately $-0.41$ and $-0.72$, respectively. The correlation functions of the 12sl state with a value larger than those of the umbrella state are $\langle \mib{\sigma}_i \cdot \mib{\sigma}_j \rangle$ described by the solid and wavy bonds, where $\mib{\sigma}_i$ and $\mib{\sigma}_j$ are A, B, C, and D spins with the positive $z$ component. It might be said that the frustration effect on $\langle \mib{\sigma}_i \cdot \mib{\sigma}_j \rangle$ is large on the solid and wavy bonds. In the mushroom state for $J/K = 2$ and $H/K = 10$, there are two types of $\langle \mib{\sigma}_i \cdot \mib{\sigma}_j \rangle$ with the strengths of approximately $0.17$ and $-0.67$, and there is only one type of $\langle h_4 \rangle$ with the strength of approximately $-1.61$. All the four-spin correlation functions of the 12sl state are larger than that of the mushroom state. Because the large absolute value of the negative correlation function indicates strong correlation, the four-spin correlations in the mushroom state are stronger than those in the 12sl state. It might be said that the frustration effect on $\langle h_4 \rangle$ is large in the 12sl state.

\begin{figure}[t]
\centering
\includegraphics[width=0.35\textwidth]{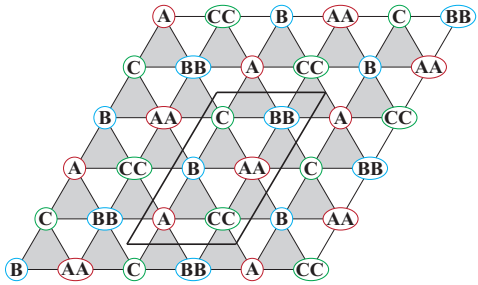}
  \caption{Example of the spin configuration of the six-sublattice structure. The marks A, B, C, AA, BB, and CC describe six types of spin orientation and the region surrounded by the solid line is the unit cell.}
\label{6sl}
\end{figure}

As the magnitude of the magnetic field is increased, the A, B, C, and D spins rotate upward, and the AA, BB, CC, and DD spins rotate downward; the system undergoes the second-order phase transition to the uud phase. As $J/K$ is decreased, the A and AA (B and BB, C and CC, D and DD) spins approach each other, and the system undergoes the first-order phase transition to the mushroom phase. On the other hand, as $J/K$ is increased, the system undergoes the first-order phase transition to the 6sl phase, as shown in Fig.~\ref{6sl}. The 6sl structure consists of six types of spin orientation, A, B, C, AA, BB, and CC. For example, when the C and CC spins have a negative $z$ component, the A, B, AA, and BB spins have a positive $z$ component.~\cite{Yasuda2007} As $J/K$ increases further, spins A and AA, B and BB, and C and CC approach each other, and the system undergoes the second-order phase transition to the phase with the umbrella structure. 

The 12sl and 6sl structures play a role in connecting the mushroom state with the four-sublattice structure stabilized by large four-spin interactions and the umbrella state with the three-sublattice structure stabilized by large two-spin interactions. In the 6sl structure with the vector chirality which does not exist in the 12sl structure, the second-order phase transition to the umbrella structure becomes possible. In the 12sl structure, the number of spins with the positive $z$ component is twice that of the spins with the negative $z$ component, and the gain in energy due to the magnetic field is more dominant than that of the 6sl structure. It would be one of the causes that expand the phase with the 12sl structure as the magnetic field is increased. Thus, the appearance of the 12sl structure conspicuously shows that all of the two- and four-spin interactions and that with the magnetic field strongly compete with each other.

\section{Phase Diagram of the System with up to  the Six-Spin Exchange Interactions}

In this section, we show the results of the multiple-spin exchange model including the five- and six-spin exchange interactions, and discuss the effects of the five- and six-spin interactions on the ground state.

\subsection{Phase diagram}

\begin{figure}[t]
  \centerline{\resizebox{0.45\textwidth}{!}{\includegraphics{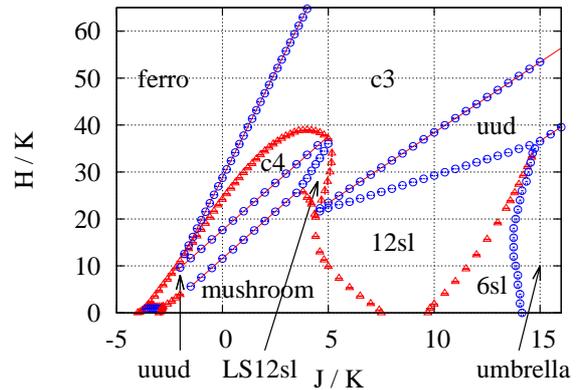}}}
  \caption{Ground-state phase diagram parametrized by $J/K$ and $H/K$ for $L/K = -0.15$ and $M/K = 0.075$, i.e., $J_5/J_4=J_6/J_4=0.3$ calculated assuming $6\times6$ sites. The triangles and circles with error bars show the first- and second-order phase transitions estimated by the CG method, respectively. The solid lines are the phase-transition lines listed in Table~\ref{hc}.}
\label{fig-phase}
\end{figure}

In Fig.~\ref{fig-phase}, we show the phase diagram of the multiple-spin exchange model including the five- and six-spin interactions. The values of $L/K$ and $M/K$ are fixed at $-0.15$ and 0.075, respectively, i.e., $J_5/J_4=J_6/J_4=0.3$~\cite{Bernu1992, Roger1998, Misguich1998, Masutomi2004}. The method explained in Sect.~3 is used to estimate the phase-transition points. Comparing Figs.~\ref{36site-phase} and \ref{fig-phase}, we find that the types of stable ground states are basically unchanged regardless of whether the five- and six-spin interactions occur. The difference between Figs.~\ref{36site-phase} and \ref{fig-phase} is that a twelve-sublattice phase appears in a region corresponding to the quadruple critical point of the mushroom, uuud, c3, and uud phases in Fig.~\ref{36site-phase}. We call the new phase the low-symmetry twelve-sublattice (LS12sl) phase, because the new phase exhibits a twelve-sublattice state with lower symmetry than that in the 12sl phase appearing in the model with up to the four-spin interactions. The single first-order phase transition between the uuud and c3 phases changes to two phase transitions owing to the appearance of the LS12sl phase. Furthermore, the phase with the u7d5 structure and a 1/6 plateau in the magnetization process appears between the phases with ferromagnetic and mushroom structures in the low-magnetic-field region. The details of these phases will be discussed in Subsects.~4.2 and 4.4, respectively.

\begin{figure}[t]
\centering
	{\resizebox{0.45\textwidth}{!}{\includegraphics{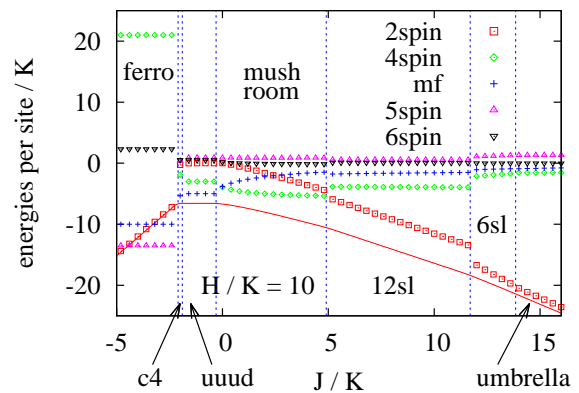}}} 
  \caption{Dependences of each energy per site on $J/K$ for $H/K = 10$ with $L/K=-0.15$ and $M/K=0.075$. The solid line shows the total energy per site. The squares, diamonds, crosses, triangles, and inverted triangles denote the energies per site of the two-spin, four-spin, magnetic-field, five-spin, and six-spin interactions, respectively. The vertical broken lines show the phase-transition points. }  
\label{fig-each_energy10}
\end{figure}

\begin{figure}[t]
\centering
	{\resizebox{0.45\textwidth}{!}{\includegraphics{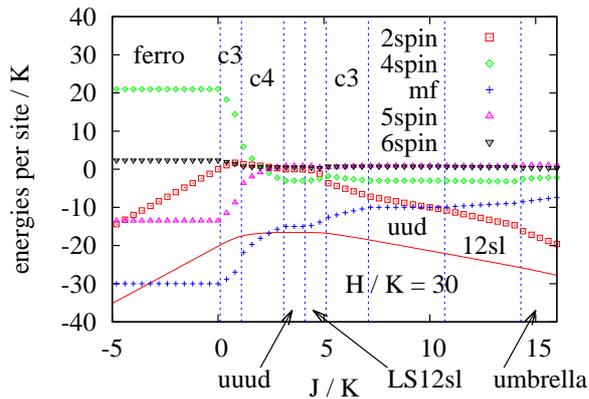}}}
  \caption{Dependences of each energy per site on $J/K$ for $H/K = 30$ with $L/K=-0.15$ and $M/K=0.075$. The lines and points denote the same as those of Fig.~\ref{fig-each_energy10}.}
\label{fig-each_energy30}
\end{figure}

The ground-state energy can be calculated as the summation of five energies of the right-hand side of Hamiltonian~(\ref{hamiltonian}). To investigate the contributions of each energy to the total energy for each phase shown in Fig.~\ref{fig-phase}, we show the $J/K$ dependences of each energy per site for $H/K = 10$ and 30 in Figs.~\ref{fig-each_energy10} and \ref{fig-each_energy30}, respectively. The solid line shows the total energy per site. The squares, diamonds, crosses, triangles, and inverted triangles denote the energies per site of the two-spin, four-spin, magnetic-field, five-spin, and six-spin interactions, respectively. 
Comparing the energies of the two- and four-spin interactions, we find that the energy of the two-spin interactions becomes small in the ferromagnetic and umbrella phases, which are on the left and right sides of the phase diagram, respectively. On the other hand, the energy of the four-spin interactions becomes small in the uuud and mushroom phases in Fig.~\ref{fig-each_energy10} and in the uuud phase in Fig.~\ref{fig-each_energy30}. As a result, there are phases where the two energies intersect between the ferromagnetic and uuud phases and between the umbrella and uuud phases. 
The energy of the magnetic field does not only decrease as $H/K$ increases, but also has a tendency to increase with $J/K$.
Although the magnitudes of the energies of the five- and six-spin interactions are small except in the ferromagnetic and c3 phases, those interactions can change the phase diagram from Fig.~\ref{36site-phase} to \ref{fig-phase}.

\begin{figure}[t]
  \centerline{\resizebox{0.45\textwidth}{!}{\includegraphics{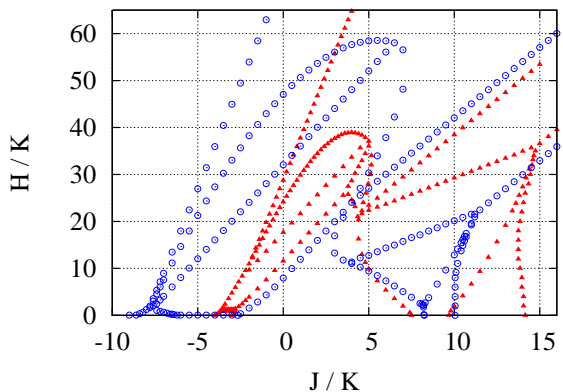}}}
  \caption{Comparison between two overlapped ground-state phase diagrams shown in Figs.~\ref{36site-phase} and \ref{fig-phase}.  The circles and solid triangles are the phase-transition points for the multiple-spin exchange model without and with the five- and six-spin interactions, respectively.}
\label{fig-phase2}
\end{figure}

Comparing the two overlapped phase diagrams in Fig.~\ref{fig-phase2}, we find that five- and six-spin exchanges expand the ferromagnetic phase. In contrast, the antiferromagnetic-like phases are contracted. This result shows that the ferromagnetic five-spin exchange interactions are dominant over the antiferromagnetic six-spin interactions, even though the interactions have the same magnitude, $J_5/J_4 = J_6/J_4 = 0.3$. The five-spin exchange is dominant over the six-spin exchange because of the difference in the number of route patterns for particle exchanges. In our model, we adopt only a regular hexagon to represent the six-spin cyclic exchange. The number of routes for the five-spin exchange is six times that for the six-spin exchange. Thus, the addition of the five- and six-spin interactions favors the ferromagnetic phase. The region of the mushroom phase undergoes little change.

\begin{figure}[t]
  \centerline{\resizebox{0.45\textwidth}{!}{\includegraphics{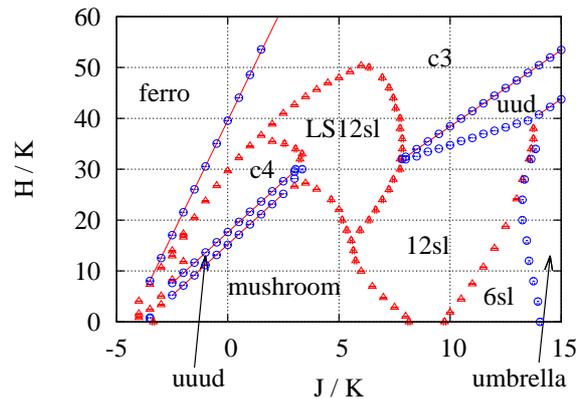}}}
  \caption{Ground-state phase diagram parametrized by $J/K$ and $H/K$ for $L/K = -0.15$ and $M/K = 0.15$, i.e., $J_5/J_4=0.3$ and $J_6/J_4=0.6$ calculated assuming $6\times6$ sites. The triangles and circles with error bars show the first- and second-order phase transitions estimated by the CG method, respectively. The solid lines are the phase-transition lines listed in Table~\ref{hc}.}
\label{fig-phase3}
\end{figure}
 
In Fig.~\ref{fig-phase3}, we show the phase diagram of the system with $L/K = -0.15$ and $M/K = 0.15$, i.e., $J_5/J_4=0.3$ and $J_6/J_4=0.6$, to examine the effects of the six-spin interactions. The region of the uud phase that extends to $J/K = 5$ in Fig.~\ref{fig-phase} shifts to a larger $J/K$ region, and the region of the LS12sl phase is expanded and is adjacent to the 12sl phase in Fig.~\ref{fig-phase3}. Although we cannot determine the order of the phase transition between the LS12sl and 12sl phases from the energies, we find a weak jump in the magnetization process. Therefore, it would be the weak first-order phase transition. Furthermore, the region of the uuud phase is contracted. 
In Fig.~\ref{fig-phase3}, a small new twelve-sublattice phase appears in the low-magnetic-field region surrounded by the uuud, c4, c3, and ferromagnetic phases.

\subsection{ Low-symmetry twelve-sublattice phase}

\begin{figure*}[t]
\centering
\includegraphics[width=0.8\textwidth]{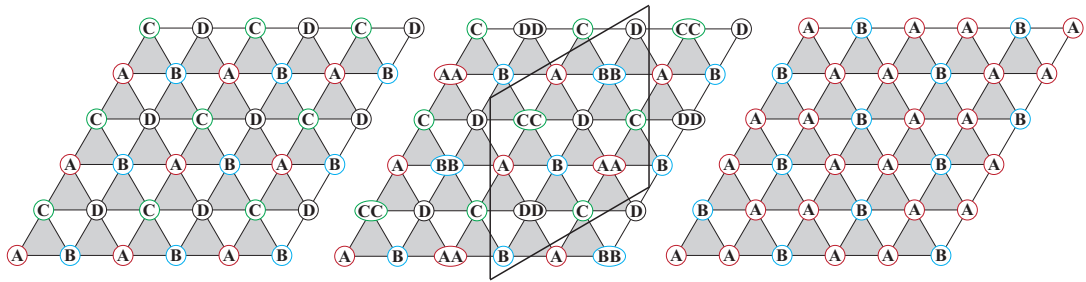}
  \caption{Example of the twelve-sublattice structure between the four- and three-sublattice structures. The left and right panels are examples of the four- and three-sublattice structures, respectively. The region surrounded by the solid line in the central panel is the unit cell of the LS12sl state. The marks denote the types of spin orientation, and A, B, C, and D in the three panels do not mean the same type of spin orientation. If the A (B, C, D) spin of the four-sublattice structure is divided into A and AA (B and BB, C and CC, D and DD) spins, the twelve-sublattice structure appears. If the A, B, C, and D (AA, BB, CC, and DD) spins in the central panel become the same spin described by A (B), the three-sublattice structure appears.}
\label{new_phase}
\end{figure*}

\begin{figure}[t]
\centering
  {\resizebox{0.4\textwidth}{!}{\includegraphics{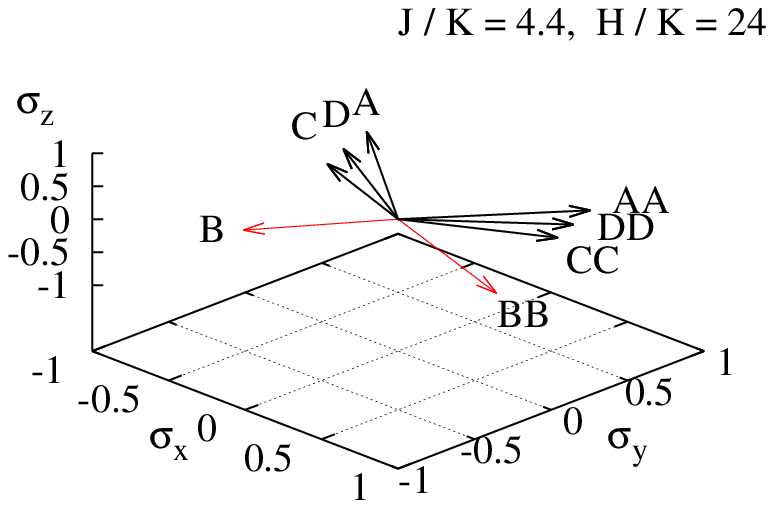}}}
  {\resizebox{0.4\textwidth}{!}{\includegraphics{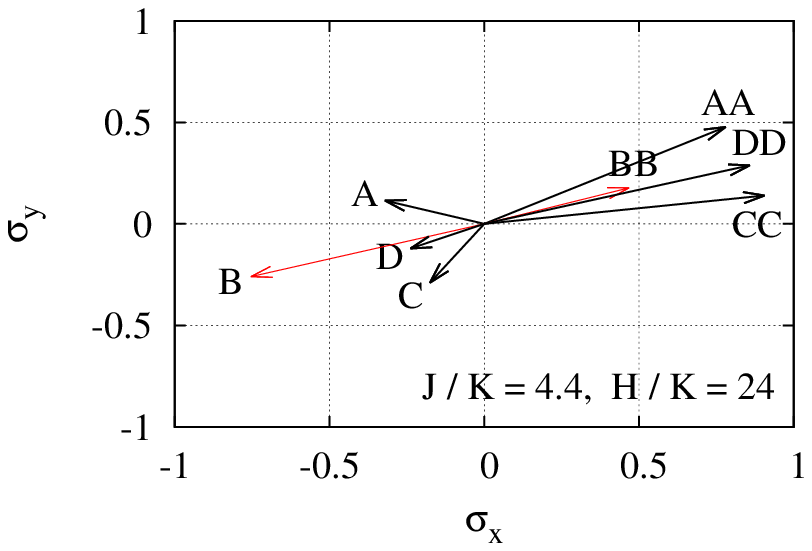}}}
  \caption{Example of eight types of spin orientation forming the twelve-sublattice structure for $J/K=4.4$, $H/K=24$, $L/K = -0.15$, and $M/K = 0.075$. The lower panel is the view from the $z$-axis direction for the upper panel. Only the B and BB spins have the negative $z$ component for this parameter. The magnitude of the spin vectors is fixed to unity.}
\label{scj44h24}
\end{figure}

The LS12sl phase, which is not observed in the system with up to the four-spin interactions, is surrounded by the mushroom, uuud, and c3 phases for $L/K = -0.15$ and $M/K = 0.075$. Because the second-order phase transition occurs between the uuud and LS12sl phases, a spin configuration that can change continuously from the uuud structure would be realized in the LS12sl phase. The phase transition between the LS12sl and c3 phases is of a weak first order. The spin configuration in the LS12sl phase is shown in the central panel of Fig.~\ref{new_phase} and is found to have a unit cell consisting of twelve sublattices. The unit cell has eight types of spin orientation, similar to that in the 12sl structure explained in Subsect.~3.3.  The eight types of spin orientation in the LS12sl phase are shown in Fig.~\ref{scj44h24}. Comparing this figure with Fig.~\ref{12sl-8spins}, we realize that the LS12sl phase has a lower-symmetry state than the 12sl phase. The left and right panels of Fig.~\ref{new_phase} illustrate the four- and three-sublattice structures, respectively. The LS12sl structure shown in the central panel of Fig.~\ref{new_phase} bridges the four-sublattice structure shown in the left panel with the three-sublattice structure shown in the right panel. When the system undergoes the first-order phase transition from the mushroom phase to the LS12sl phase, each of the four types of spin orientation in the mushroom phase is divided into two types of spin orientation. For example, the A spin in the left panel of Fig.~\ref{new_phase} is divided into the A and AA spins described in the central panel of Fig.~\ref{new_phase}. The spin state in the LS12sl phase approaches a coplanar state as the system approaches the c3 phase. When the system undergoes the phase transition to the c3 phase, the orientations of the A, B, C, and D spins become the same and those of the AA, BB, CC, and DD spins also become the same simultaneously. On the other hand, as the system approaches the uuud phase from the LS12sl phase, the spin state in the LS12sl phase approaches the collinear state. When the system undergoes the second-order phase transition to the uuud phase, the orientations of the A and AA (B and BB, C and CC, D and DD) spins become the same.

\subsection{Magnetization process}

\begin{figure}[t]
  \centerline{\resizebox{0.45\textwidth}{!}{\includegraphics{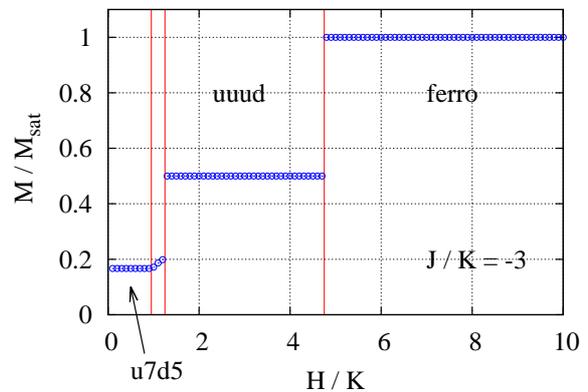}}}
  \caption{Magnetization process for $J/K=-3$, $L/K = -0.15$, and $M/K = 0.075$. The system has the 1/6, 1/2, and 1 plateaus. The solid lines show the phase-transition points in the finite magnetic field.}
\label{magpro-3}
\end{figure}

In this subsection, we discuss the magnetization process for $L/K = -0.15$ and $M/K = 0.075$. We show the magnetization process for $J/K = -3$ in Fig.~\ref{magpro-3}. The horizontal and vertical axes represent the magnitude of the magnetic field and the magnetization per $M_{\rm sat}$, respectively, where  $M_{\rm sat}$ is the saturated value of the magnetization. In the magnetization process, a 1/6 plateau appears at a low magnetic field. The spin configuration with the 1/6 plateau has the 12-sublattice structure, in which the unit cell consists of seven spins parallel to and five spins antiparallel to the magnetic field. We call it the u7d5 phase and will discuss the details in the next subsection. Because the 1/6 plateau does not appear for the system without the five- and six-spin interactions, the 1/6 plateau occurs as a result of the five- and six-spin interactions. The ground state at $H/K=0$ is a degenerate state with various magnetizations. The phase between the u7d5 and uuud phases is the canted u7d5 phase. As the magnitude of the magnetic field is increased, the system undergoes the first-order phase transition twice and 
 reaches the ferromagnetic phase.

\begin{figure}[t]
  \centerline{\resizebox{0.45\textwidth}{!}{\includegraphics{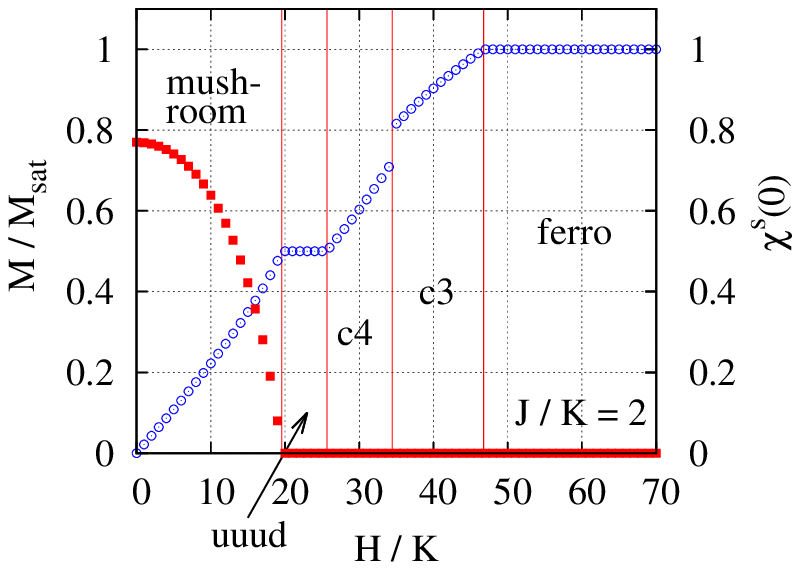}}}
  \caption{Magnetization process and dependence of the scalar chirality $\chi^{\rm s}(\mib{0})$ on the magnetic field for $J/K=2$, $L/K = -0.15$, and $M/K = 0.075$. The left and right $y$-axes denote the magnetization (circles) and $\chi^{\rm s}(\mib{0})$ (solid squares), respectively. The solid lines show the phase-transition points.}
\label{magpro2}
\end{figure}

\begin{figure}[t]
  \centerline{\resizebox{0.45\textwidth}{!}{\includegraphics{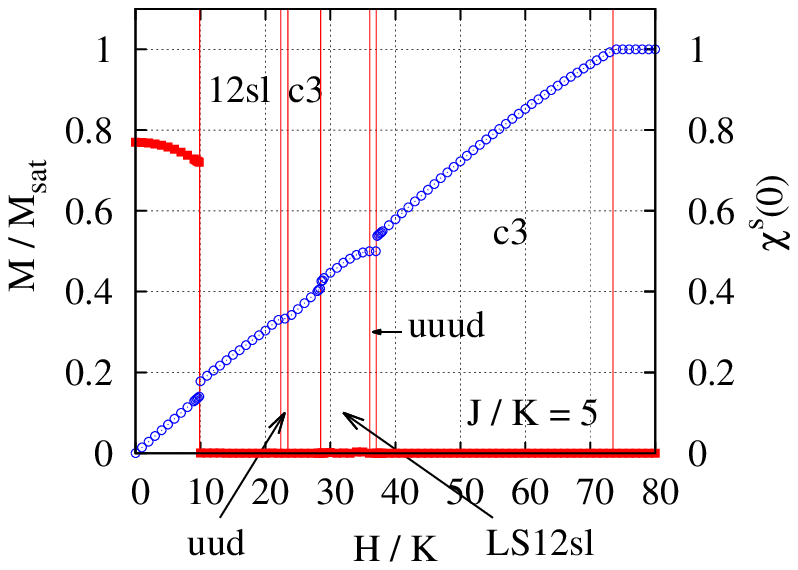}}}
  \caption{Magnetization process and dependence of the scalar chirality $\chi^{\rm s}(\mib{0})$ on the magnetic field for $J/K=5$, $L/K = -0.15$, and $M/K = 0.075$. The left and right $y$-axes denote the magnetization (circles) and $\chi^{\rm s}(\mib{0})$ (solid squares), respectively. The solid lines show the phase-transition points.}
\label{magpro5}
\end{figure}

The magnetization process and the dependence of the scalar chirality $\chi^{\rm s}(\mib{0})$ on the magnetic field for $J/K=2$ are shown in Fig.~\ref{magpro2}. The circles and solid squares denote the magnetization and $\chi^{\rm s}(\mib{0})$, respectively. For $0 \le H/K < 20$, the finite $\chi^{\rm s}(\mib{0})$ shows that the system is in the mushroom phase. For $H/K \ge 20$, the regions with the plateaus $M/M_{\rm sat}=0.5$ and 1 are the uuud and ferromagnetic phases, respectively. A first-order phase transition with a jump in the magnetization exists between the c4 and c3 phases. The magnetization process and the dependence of the scalar chirality $\chi^{\rm s}(\mib{0})$ on the magnetic field are shown in Fig.~\ref{magpro5} for $J/K=5$, where the system realizes the most varied phases. For $0 \le H/K < 10$, the mushroom phase with finite $\chi^{\rm s}(\mib{0})$ appears. A first-order phase transition with a jump in the magnetization exists between the mushroom and 12sl phases. The c3 and LS12sl phases appear between the uud and uuud phases. As the magnetic field is increased, the system passes through the c3 phase and reaches the ferromagnetic phase.

\begin{figure}[t]
  \centerline{\resizebox{0.45\textwidth}{!}{\includegraphics{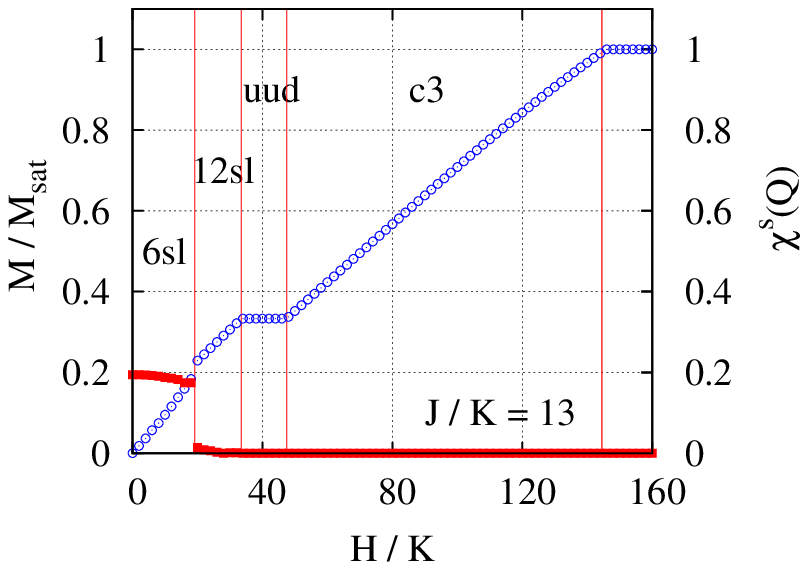}}}
  \caption{Magnetization process and dependence of the staggered scalar chirality $\chi^{\rm s}(\mib{Q})$ on the magnetic field for $J/K=13$, $L/K = -0.15$, and $M/K = 0.075$. The left and right $y$-axes denote the magnetization (circles) and $\chi^{\rm s}(\mib{Q})$ (solid squares), respectively. The solid lines show the phase-transition points.}
\label{magpro13}
\end{figure}

\begin{figure}[t]
  \centerline{\resizebox{0.45\textwidth}{!}{\includegraphics{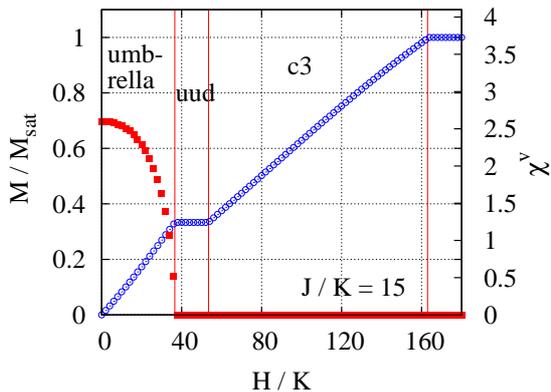}}}
  \caption{Magnetization process and dependence of the vector chirality $\chi^{\rm v}$ on the magnetic field for $J/K=15$, $L/K = -0.15$, and $M/K = 0.075$. The left and right $y$-axes denote the magnetization (circles) and $\chi^{\rm v}$ (solid squares), respectively. The solid lines show the phase-transition points.}
\label{magpro15}
\end{figure}

We show the magnetization process and $H/K$ dependence of the staggered scalar chirality $\chi^{\rm s}(\mib{Q})$ for $J/K = 13$ in Fig.~\ref{magpro13}. The largest value among the observed $\chi^{\rm s}(\mib{Q})$ with $\mib{Q}=(\pi,\pi)$, $(\pi,0)$, and $(0,\pi)$ is plotted as the value of $\chi^{\rm s}(\mib{Q})$ in Fig.~\ref{magpro13}. The 6sl structure with finite $\chi^{\rm s}(\mib{Q})$ exists for $0 \le H/K < 20$, and the first-order phase transition to the 12sl structure occurs. For $H/K \ge 20$, the uud phase with $M/M_{\rm sat}=1/3$ and the ferromagnetic phase with $M/M_{\rm sat}=1$ appear. Finally, we show the magnetization process and the $H/K$ dependence of the vector chirality $\chi^{\rm v} = |\mib{\chi}^{\rm v}|$ for $J/K = 15$ in Fig.~\ref{magpro15}. The umbrella structure with finite $\chi^{\rm v}$ appears for $0 \le H/K < 37$. As the magnetic field is increased, the system passes through the uud and c3 phases and reaches the ferromagnetic phase.

\subsection{u7d5 phase}

\begin{figure}[t]
  \centerline{\resizebox{0.45\textwidth}{!}{\includegraphics{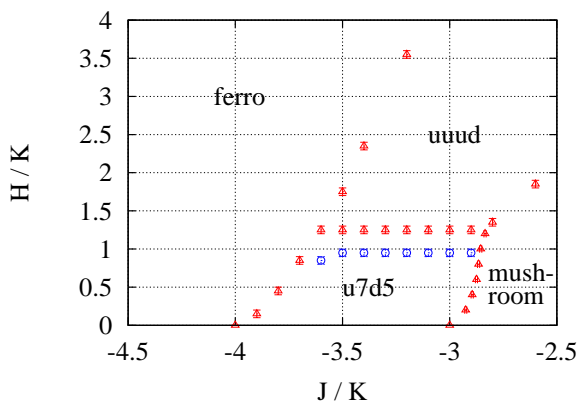}}}
  \caption{Ground-state phase diagram parametrized by $J/K$ and $H/K$ for $L/K = -0.15$ and $M/K = 0.075$, i.e., $J_5/J_4=J_6/J_4=0.3$. The triangles and circles with error bars denote the first- and second-order phase transitions, respectively. The phase between the u7d5 and uuud phases is the canted u7d5 phase.}
\label{u7d5phase1}
\end{figure}

\begin{figure}[t]
\centering
\includegraphics[width=0.35\textwidth]{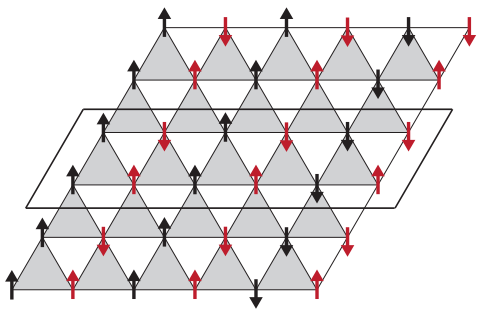}
  \caption{Example of the spin configuration of the u7d5 structure. The region surrounded by the solid line is the unit cell.}
\label{u7d5}
\end{figure}

When we take the five- and six-spin interactions into account, the u7d5 phase appears for $-4 < J/K < -2.8$ and $0 < H/K < 1$ with $L/K = -0.15$, and $M/K = 0.075$. In Fig.~\ref{u7d5phase1}, we show an enlarged view of Fig.~\ref{fig-phase}. A magnetization plateau with $M/M_{\rm sat}=1/6$ exists in the u7d5 phase, as shown in the magnetization process in Fig.~\ref{magpro-3}. The u7d5 phase does not appear at $H/K=0$ and is induced by the magnetic field. For $-4 < J/K \le -3.7$, the system undergoes the first-order phase transition from the u7d5 phase to the ferromagnetic phase as the magnetic field is increased. On the other hand, for $-3.6 \le J/K \le -2.9$, after the system undergoes the second-order phase transition to a phase with a magnetization $M/M_{\rm sat} > 1/6$, the first-order phase transition to the uuud phase occurs. The spins of the phase between the u7d5 and uuud phases are canted with respect to the magnetic field. An example of the spin configuration and the sublattice structure of the u7d5 state is shown in Fig.~\ref{u7d5}. The spin configuration is regarded as coupled one-dimensional antiferromagnetic and ferromagnetic chains. The antiferromagnetic and ferromagnetic chains are coupled alternately, and the spin configurations of the antiferromagnetic chains are in phase. The unit cell of the u7d5 state is shown in Fig.~\ref{u7d5}. Parts of three ferromagnetic chains are included in the unit cell, and the spins of two of the parts are parallel to the magnetic field.

\begin{figure}[t]
  \centerline{\resizebox{0.45\textwidth}{!}{\includegraphics{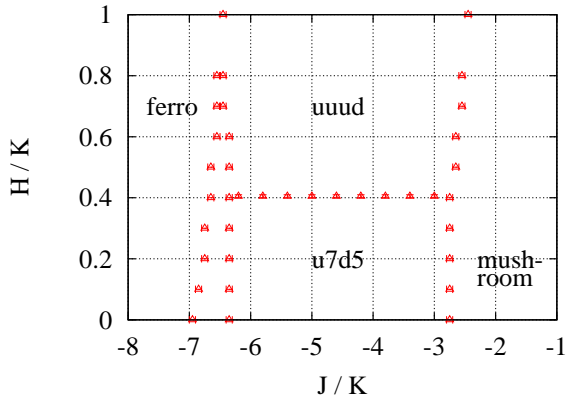}}}
  \caption{Ground-state phase diagram parametrized by $J/K$ and $H/K$ for $L/K = -0.05$ and $M/K = 0.025$, i.e., $J_5/J_4=J_6/J_4=0.1$. The triangles denote the first-order phase transitions. In the region surrounded by the u7d5, uuud, and ferromagnetic phases, there are small phases with various magnetizations.}
\label{u7d5phase2}
\end{figure}

\begin{figure}[t]
  \centerline{\resizebox{0.45\textwidth}{!}{\includegraphics{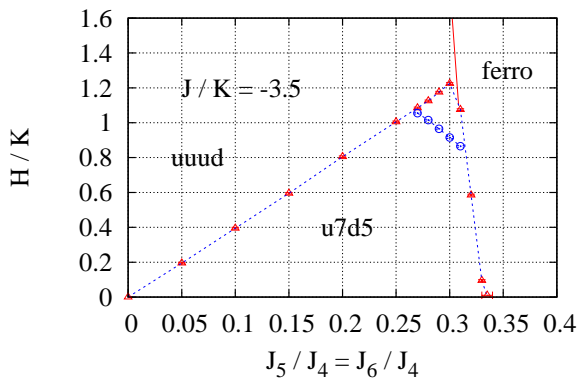}}}
 \caption{Ground-state phase diagram parametrized by $J_5/J_4=J_6/J_4$ and $H/K$ for $J/K = -3.5$. The triangles and circles denote the first- and second-order phase transitions, respectively. The phase surrounded by the u7d5, uuud, and ferromagnetic phases is the canted u7d5 phase. The solid line is the phase-transition line between the uuud and ferromagnetic phases calculated analytically. The broken lines are guides to the eyes.}
\label{u7d5phase3}
\end{figure}

To investigate the effects of the five- and six-spin interactions on the u7d5 phase, we show the phase diagram for $L/K=-0.05$ and $M/K=0.025$, i.e., $J_5/J_4=J_6/J_4=0.1$ in Fig.~\ref{u7d5phase2}. The shape of the u7d5 phase is contracted along the $H/K$ axis and expanded along the $J/K$ axis in comparison with that for $L/K=-0.15$ and $M/K=0.075$. In a triangular region $-7 < J/K < -6.3$ surrounded by the u7d5, uuud, and ferromagnetic phases, there are small phases with various structures depending on $J/K$. The phase diagram parameterized by $J_5/J_4=J_6/J_4$ and $H/K$ for $J/K = -3.5$ is shown in Fig.~\ref{u7d5phase3}. The region of the u7d5 phase, which does not exist in the system without the five- and six-spin interactions, is expanded with $J_5/J_4$ and shrinks rapidly near $J_5/J_4 = 0.3$. The phase surrounded by the u7d5, uuud, and ferromagnetic phases is the canted u7d5 phase.

\begin{figure}[t]
  \centerline{\resizebox{0.45\textwidth}{!}{\includegraphics{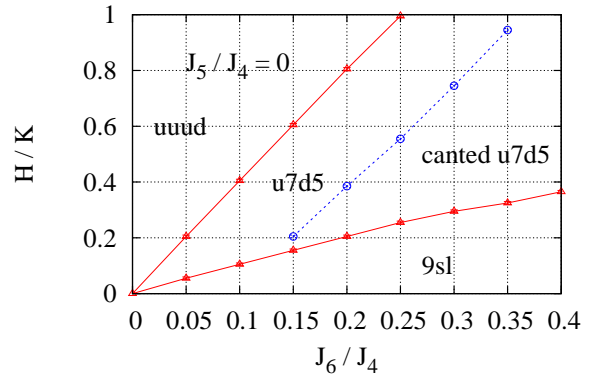}}}
  \caption{Ground-state phase diagram parametrized by $J_6/J_4$ and $H/K$ for $L/K = 0$ ($J_5/J_4 = 0$) and $J/K = -3.5$. The triangles and circles denote the first- and second-order phase transitions, respectively. The lines are guides to the eyes.}
\label{u7d5phase4}
\end{figure}

Although the u7d5 phase does not appear for finite $J_5$ and $J_6=0$, it appears for $J_5=0$ and finite $J_6$, as shown in Fig.~\ref{u7d5phase4}. Thus, it is found that the contribution of the six-spin interactions to the appearance of the u7d5 phase is dominant. For $J_5=0$, a canted phase of the u7d5 phase and the phase with the nine-sublattice structure appear in the low-field region of the u7d5 phase. The unit cell of the nine-sublattice structure consists of three spins each for the three types of spin orientation. The three types of spin orientation are similar to those constituting the umbrella structure close to the 120$^{\circ}$ structure. Therefore, the magnetization of the phase is almost zero, and the ground-state energy hardly varies with the magnetic field. Because the phase with the nine-sublattice structure does not appear for $J_5/J_4=J_6/J_4$, it is concluded that the five-spin interactions play a role in destroying this phase.

\section{Summary and Discussion}

In this work, we investigated the multiple-spin exchange model with up to the six-spin interactions. We calculated the ground-state energy assuming a $6\times6$ sublattice within the mean-field approximation and determined the ground-state phase diagram parameterized by the interaction parameter $J/K$ and the magnetic field $H/K$. 
 
In the model with up to the four-spin interactions, we determined the order of the phase transition. As a result, we found that the order of the phase transition between the mushroom and uuud phases depends on $J/K$. Furthermore, we investigated the precise twelve-sublattice structure surrounded by the phases with mushroom, uud, and six-sublattice structures.
 
Comparing the three phase diagrams for $(L/K,~M/K)=(0,~0)$, $(-0.15,~0.075)$, and $(-0.15,~0.15)$, we found that the five- and six-spin exchange expands the ferromagnetic phase. In contrast, the antiferromagnetic-like phases are contracted. This trend occurs because there are more five-spin interactions than six-spin interactions. Furthermore, we found two new phases induced by the five- and six-spin interactions. One is the LS12sl phase, which bridges the mushroom and c3 phases. The region of the 12sl and LS12sl phases is expanded by the six-spin interactions. The other is the u7d5 phase, which is stabilized in the low-magnetic-field region under the uuud phase. The unit cell of this phase, like that of the uud and uuud phases, consists of up and down spins, and a 1/6 plateau appears in the magnetization process. Although the region of the u7d5 phase is expanded by the six-spin exchange, the u7d5 phase becomes unstable under the five-spin exchange.

In this work, we found that the critical magnetic field $H_{\rm sat}$ at which the system is in the ferromagnetic phase becomes small owing to the five- and six-spin interactions and that the uuud phase with the 1/2 plateau is contracted by the five- and six-spin interactions. Nema {\it et al.} reported the magnetization curve of a $^3$He film adsorbed on a graphite surface~\cite{Nema2009} and pointed out that $H_{\rm sat}$ evaluated on the basis of the mean-field theory and exact diagonalization on finite-size systems for the model with up to the four-spin interactions~\cite{Momoi1999} is larger than the experimental value, and that the magnetic field region corresponding to the 1/2 plateau evaluated on the basis of the theoretical work is wider than the experimental result. The results of the present work showed the importance of the five- and six-spin interactions in the construction of a theoretical model that can explain the experimental results quantitatively.

The ground state of the multiple-spin exchange model with up to the six-spin interactions has been investigated by the exact diagonalization of a 24-site quantum system.~\cite{Misguich1998} In that work, the 1/2 plateau in the magnetization process appears for $J/K < 0$, and it is predicted that the ground state of the lower-magnetic-field region below the 1/2 plateau phase is a nonmagnetic gapped spin-liquid phase. The 1/6 plateau phase found in the present work has not been found in finite quantum systems. Because the classical description becomes adequate at a high magnetic field, we can understand that the 1/2 plateau obtained in the classical system survives quantum fluctuations. On the other hand, the 1/6 plateau phase that appears in a weak field region might be fragile against quantum fluctuations. In addition, there is a possibility that the appearance of the 1/6 plateau might be due to a finite-size effect of this calculation as observed in the ANNNI model.~\cite{Selke1988} Therefore, future studies of the instability of such a 1/6 plateau against quantum fluctuations and size effects are desired. An experimental discovery of the 1/6 plateau for real compounds would be useful for estimating the exchange interactions, $J_4$, $J_5$, and $J_6$.

For $J/K > 0$ in the multiple-spin exchange model with up to the four-spin interactions, a 1/3 plateau has been found in 27-site quantum systems.~\cite{LiMing2000} This 1/3 plateau also appears within the mean-field approximation.~\cite{Kubo1997} LiMing {\it et al}. concluded that the ground state in the low-magnetic-field region for $J/K > 0$ is a gapless quantum spin liquid.~\cite{LiMing2000} On the other hand, the ground state obtained within the mean-field approximation for the same parameter region is a six- or twelve-sublattice structure with a chiral order. The effects of quantum fluctuations on the ordered phase with the short-range structure of this chirality are known to be weak~\cite{Momoi1992}. It would be interesting to study whether the chiral order obtained within the mean-field approximation in this work can survive quantum fluctuations.

\section*{Acknowledgment}
This work was supported by JSPS KAKENHI Grant Number JP16K05479.

\end{document}